\documentstyle[12pt]{article}
\newif\ifprelim
\def\mode{b }
\def\unlock{\catcode`@=11 }% This allows us to modify PLAIN macros.
\def\lock{\catcode`@=12 }% at signs are no longer letters

\def\marginnote#1{}

\def\Newif#1{\expandafter\ifx\csname#1\endcsname\relax
 \csname newif\expandafter\endcsname\csname#1\endcsname\fi}
\Newif{ifprelim}

\ifx\mode\undefined\read-1 to\mode\fi
\unlock
\def\eqnlabels{
        \def\draftlabel##1{{\@bsphack\if@filesw {\let\thepage\relax
           \xdef\@gtempa{\write\@auxout{\string
              \newlabel{##1}{{\@currentlabel}{\thepage}}}}}\@gtempa
           \if@nobreak \ifvmode\nobreak\fi\fi\fi\@esphack}
                \gdef\@eqnlabel{##1}}
        \def\@eqnlabel{}
        \def\@vacuum{}
        \let\label=\draftlabel
        \def\@eqnnum{(\theequation)\rlap{\kern\marginparsep\tt\@eqnlabel}%
                \global\let\@eqnlabel\@vacuum}
}
\lock

\unlock
\newif\if@defeqnsw \@defeqnswtrue

\def\eqnarray{\stepcounter{equation}\let\@currentlabel=\theequation
\if@defeqnsw\global\@eqnswtrue\else\global\@eqnswfalse\fi
\global\@eqnswtrue
\tabskip\@centering\let\\=\@eqncr
$$\halign to \displaywidth\bgroup\hfil\global\@eqcnt\z@
  $\displaystyle\tabskip\z@{##}$&\global\@eqcnt\@ne
  \hfil$\displaystyle{{}##{}}$\hfil
  &\global\@eqcnt\tw@ $\displaystyle{##}$\hfil
  \tabskip\@centering&\llap{##}\tabskip\z@\cr}

\def\yesnumber{\global\@eqnswtrue}

\def\@@eqncr{\let\@tempa\relax\global\advance\@eqcnt by \@ne
    \ifcase\@eqcnt \def\@tempa{& & & &}\or \def\@tempa{& & &}\or
     \def\@tempa{& &}\or \def\@tempa{&}\else\fi
     \@tempa \if@eqnsw\@eqnnum\stepcounter{equation}\fi
     \if@defeqnsw\global\@eqnswtrue\else\global\@eqnswfalse\fi
     \global\@eqcnt\z@\cr}

\def\@eqnacr{{\ifnum0=`}\fi\@ifstar{\@yeqnacr}{\@yeqnacr}}

\def\@yeqnacr{\@ifnextchar [{\@xeqnacr}{\@xeqnacr[\z@]}}

\def\@xeqnacr[#1]{\ifnum0=`{\fi}\cr \noalign{\vskip\jot\vskip #1\relax}}

\def\eqalign{\null\,\vcenter\bgroup\openup1\jot \m@th \let\\=\@eqnacr
\ialign\bgroup\strut
\hfil$\displaystyle{##}$&$\displaystyle{{}##}$\hfil\crcr}
\def\endeqalign{\crcr\egroup\egroup\,}

\def\eqalignno{\stepcounter{equation}\let\@currentlabel=\theequation
\if@defeqnsw\global\@eqnswtrue\else\global\@eqnswfalse\fi
\let\\=\@eqncr
$$\displ@y \tabskip\@centering \halign to \displaywidth\bgroup
  \global\@eqcnt\@ne\hfil
  $\@lign\displaystyle{##}$\tabskip\z@skip&\global\@eqcnt\tw@
  $\@lign\displaystyle{{}##}$\hfil\tabskip\@centering&
  \llap{\@lign##}\tabskip\z@skip\crcr}
\def\endeqalignno{\@@eqncr\egroup
      \global\advance\c@equation\m@ne$$\global\@ignoretrue}
\lock

\catcode`\@=11
\font\tenmsb=msbm10
\font\sevenmsb=msbm7 \font\fivemsb=msbm5 \newfam\msafam
\newfam\msbfam
\textfont\msbfam=\tenmsb
\scriptfont\msbfam=\sevenmsb \scriptscriptfont\msbfam=\fivemsb
\def\hexnumber@#1{\ifnum#1<10 \number#1\else \ifnum#1=10
A\else\ifnum#1=11
 B\else\ifnum#1=12 C\else \ifnum#1=13 D\else\ifnum#1=14
E\else\ifnum#1=15
 F\fi\fi\fi\fi\fi\fi\fi}
\def\msa@{\hexnumber@\msafam} \def\msb@{\hexnumber@\msbfam}
\def\Bbb{\ifmmode\let\next\Bbb@\else
\def\next{\errmessage{Use \string\Bbb\space only in math
mode}}\fi\next}
\def\Bbb@#1{{\Bbb@@{#1}}} \def\Bbb@@#1{\fam\msbfam#1}
\mathchardef\varkappa="0\msb@7B
\catcode`\@=12

\def\bm#1{\mbox{\boldmath $#1$}}
\newcommand{\sect}[1]{Sect.\ref{#1}}
\catcode `\@=11
\@addtoreset{equation}{section}
\def\theequation{\arabic{section}.\arabic{equation}}
\catcode `\@=12
\date{}
\title{Group theoretical examination of the \\ relativistic wave equations on
curved spaces.\\ III. Real reducible spaces}
\author{S.A.Pol'shin }
\begin{document}
\maketitle
\thispagestyle{empty}
\begin{abstract}
The group theoretical approach to the relativistic wave equations
on the real reducible
 spaces  for spin~0, 1/2 and~1 massless particles  is
considered. The invariant wave equations which determine the appropriate
irreducible representations are constructed.  The coincidence of these
equations with the general-covariant Klein-Gordon, Weyl and Maxwell equations
on the corresponding spaces is shown.
The explicit solutions of these equations possessing a simplicity and
physical transparency are obtained in the form of so-called "plane waves"
without using  the method of separation of variables.
The invariance properties of these "plane waves" for the
spinless particles under the group $SO(3,1)$  were used for the
construction of the invariant spin~0,1/2 and~1 two-point functions on the
$H^3$.  Secondly quantized spin~0,1/2 and~1 fields on the
${\Bbb R}^{1}\otimes H^{3}$ are constructed; their propagators which are
their (anti)commutators in  different points, are expressed in terms of the
mentioned two-point functions.  From here the  ${\Bbb R}^{1}\otimes
 SO(3,1)$-invariance  follows.
\end{abstract}

\section{Introduction}
In the present paper we
 consider the relativistic wave equations for the massless particles in real
 reducible spaces following the program outlined in~\cite{I}.
The Einstein space ${\Bbb R}^1 \otimes S^3$ is the most
important of them; the wave equations on this space were
 considered in a number of papers by M.Carmeli with co-authors using
  some group-theoretical methods (see~\cite{42}). We will devote the main
 part of our paper to the space ${\Bbb R}^1 \otimes H^3$ which differs
 from the space ${\Bbb R}^1 \otimes S^3$ only by the sign of curvature. Our
 choice  based on  two important advantages of the space ${\Bbb R}^1 \otimes
 H^3$ over the space ${\Bbb R}^1 \otimes S^3$. The first advantage consists in
 that the symmetry group of the space ${\Bbb R}^1 \otimes H^3$ has
 infinite-dimensional unitary irreducible representations; the second
one consists in that the solutions of the group-theoretical wave equations on
 the space ${\Bbb R}^1 \otimes H^3$ possess the remarkable invariance
 properties under the symmetry group of the space, which allows us to
 construct  invariant propagators with the help of the mentioned solutions.

 The present paper is constructed as follows.
In~\sect{III.1} the space $H^3$, its symmetry group and its irreducible
representations are considered. \sect{III.2} is devoted to the massless spin
zero particles on the ${\Bbb R}^1 \otimes H^3$. It is shown that the
corresponding invariant wave operator coincides with the covariant
d'Alembertian; the analogous situation takes place  in the de Sitter
space~\cite{II} too. \sect{III.3} treats the wave equations for the massless
spin 1/2 and 1 fields. In~\sect{III.4} we consider the exact solutions of
these equations in the form of so-called "plane waves". In~\sect{III.5}  is
shown that the obtained group-theoretical wave equations for the spin 1/2 and
1 particles coincides with the general-covariant Weyl and Maxwell equations
on the ${\Bbb R}^1 \otimes H^3$.

In~\sect{III.6} we apply the obtained exact solutions of  wave equations
to the construction of the secondly-quantized spin 0,~1/2 and~1 fields on the
${\Bbb R}^1 \otimes H^3$ and their invariant propagators.
The obtaining of  invariant propagators is an important problem
of the quantum field theory on the curved spaces. Earlier the propagators of
 various spin fields on the de Sitter and Anti-de Sitter spaces
(see~\cite{caus4,caus5,caus6} and references therein)
and spin 0 and 1/2 fields on the $S^n$~\cite{45} were considered. In the
recent paper~\cite{43} the propagators for massive spin 0 and 1/2 fields over
the ${\Bbb R}^{1}\otimes S^{n}$ space are constructed. In the above referred
 papers the propagators are constructed as  invariant functions obeys the
corresponding wave equations. In this case its connection with the
(anti)commutators of  secondly quantized fields is unclear.  In
the~\sect{III.6} we start from the invariance properties of the obtained
"plane waves" ,which properties based on the connection of these "plane
waves" with the generalized coherent states for the group
$SO(3,1)$. Using these properties we construct the invariant
spin 0,1/2 and~1 two-point functions on the $H^3$.
The propagators of these fields
 being their (anti)commutators in the different points, are expressed
through the mentioned two-point functions and possess the invariance
under the symmetry group of our space.

In~\sect{III.7} we consider the wave equations on the other reducible spaces:
Einstein space and one more space with the symmetry group $T_{1}^{\Bbb
R}\otimes SO(3)$.

Indices $\alpha,\beta,\ldots$
correspond to the embracing space for $H^3$ and run values 1 up to 4;
indices $\mu,\nu,\ldots$ correspond to the space ${\Bbb R}^{1}\otimes
H^3$ and run values 0 up to 3. Indices $i,k,\ldots=1,2,3$.

\section{$H^3$ space and its symmetry group \label{III.1}}
The $H^{3}$ space is the hypersphere in the fictitious four-dimensional
pseudo-euclidean space with the metric $\eta_{\alpha\beta}=
\mbox{diag }(+1,+1,+1,-1)$; this hypersphere is determined by the equation
\begin{equation}\label{4.  1}
\eta_{\alpha\beta}x^{\alpha}x^{\beta}=-R^{2}.
\end{equation}
 As the interval looks like $ds^{2}=dx^{\alpha}dx_{\alpha}, $
then, by expressing $x^4$ from~(\ref{4.  1}):
$$x^{4}=\varkappa\equiv\sqrt{1+{\bf x}^{2}/R^{2}},$$
 we obtain
\begin{equation}\label{4.  2}
\begin{eqalign}
ds^{2}=g_{ij}dx^{i}dx^{j} \\
 g_{ij}=\delta_{ij}-\frac{x^{i}x^{j}}{R^{2}\varkappa ^{2}} \\
g^{ij}=\delta^{ij}+\frac{x^{i}x^{j}}{R^{2}}.
\end{eqalign}
\end{equation}
By the choice of $x^4$ sign, we have fixed one of the halves of the space,
which is covered by our coordinates.  The symmetry group of the space is,
obviously, $SO(3,1)$ with generators of rotations $\bf X$ and translations
$\bf P$ and commutation relations
 \begin{eqalignno}\label{4.  3/1}
{[}P_{i}, P_{j}{]}=-R^{-2}\varepsilon_{ijk}X_{k} \\
 {[}X_{i}, X_{j}{]}=\varepsilon_{ijk}X_{k} \label{4. 3/2}      \\
 {[}P_{i}, X_{j}{]}=\varepsilon_{ijk}P_{k}. \label{4. 3/3}
 \end{eqalignno}
With $R \rightarrow\infty$  these commutation relations turn into
commutation relations of the three-dimensional Euclidean group.

Our  group has two second-order Casimir operators:
$$\begin{eqalign}
C_{2}^{(1)}=P_{i}P_{i}-R^{-2}X_{i}X_{i} \\
  C_{2}^{(2)}=P_{i}X_{i}.
\end{eqalign}$$
It is easy to show that the generators
\begin{equation}\label{4. 3a}
 \Pi ^{\pm}_{i}=P_{i} \pm iR^{-1}X_{i}
 \end{equation}
 compose two $SO(3)$-like subgroups commuting with each other:
$$\begin{eqalign}
  {[}\Pi ^{\pm}_{i},
\Pi^{\pm}_{j}{]}=\pm \frac{2i}{R}\varepsilon _{ijk}\Pi ^{\pm}_{k}; \\
 {[}\Pi ^{+}_{i}, \Pi ^{-}_{j} {]}=0.
\end{eqalign}$$
Since the presence of the imaginary unit in~(\ref{4. 3a}) the isomorphism of
the $SO(3,1)$ and $SO(3)\otimes SO(3)$ takes place only at a level of the Lie
algebras. The Casimir operators of these subgroups
\begin{equation}\label{4. 7}
-R^{-2} C_{2}^{\pm}=\frac{1}{4}\Pi_{i}^{\pm}\Pi_{i}^{\pm} =
\frac{1}{4} C_{2}^{(1)}\pm  \frac{1}{2} iR^{-1} C^{(2)}_{2}
\end{equation}
are the Casimir operators of the all group. 
The irreducible representations
of the $SO(3,1)$ group are well-known~\cite{49}; they are determined by
two (complex, in general) numbers
 $j_+$ and $j_-$ such as
\begin{equation}\label{4. 8}
j_{+}-j_{-}=\pm  s, \quad s=0, 1/2, 1,\ldots
\end{equation}
From the Weinberg theorem~\cite{48} it  follows  that such a
representation should describe the spin $s$ particles.
If $j_+ +j_{-} $ also is an integer or half-integer  then the
representation is finite-dimensional and non-unitary; in the opposite case
 the representation is infinite-dimensional and unitary.
 In the representation ${\sf L}^{(j_{+},j_- )}$ the eigenvalues of the Casimir
 operators are
 \begin{equation}\label{4. 8a}
 C_{2}^{\pm}=-j_{\pm}(j_{\pm}+1).
\end{equation}
As the operators $C_{2}^{(1)}$ and $C_{2}^{(2)}$ are Hermitian then
 those eigenvalues should be real; from here follows
 $C_{2}^{+}=\overline{C}
\vphantom{C}^{-}_{2}$; combining with~(\ref{4. 8a}) one can obtain
$$(\overline{\jmath}_{+}
-j_{-})(\overline{\jmath}_{+}+j_{-}+1)=0.$$
If the first bracket is equal to zero then we obtain  the additional
 series of  representations which is not interest for us. If the
 second bracket is equal to zero then we came to the principal series of
 representations and from~(\ref{4. 8}) follows
\begin{equation}\label{4. 8b}
\mbox{Re}\, j_+ =\frac{\pm s-1}{2} \quad,\quad
\mbox{Re}\, j_- =\frac{\mp s-1}{2}.
\end{equation}
The metric on the ${\Bbb R}^{1}\otimes H^3$ is determined
 by the formulas~(\ref{4.  2}) and $g_{00}=-1,\ g_{0i}=0.$

\section{Spinless particles and Klein-Gordon equation  \label{III.2}}

As for the metric~(\ref{4.  2})
the Christoffel symbols, not equal to zero, are
$$\Gamma ^{k}_{ij}=R^{-2}g_{ij}x^{k}, $$
then it is easy to obtain the Killing vectors,  which  give us the generators
of scalar representation:
\begin{equation}\label{4.  5}
P^{(l)}_{i}=\varkappa \partial _{i} \quad,\quad
X^{(l)}_{i}=\varepsilon _{ijk} x^{k}\partial _{j}.
\end{equation}
Therefore, the generators of  scalar representation
of both $SO(3)$-subgroups  looks like
\begin{equation}\label{4. 5a}
 \Pi^{+(l)}_{i}=e^{k}_{(i)}\partial _{k} \qquad,\qquad
 \Pi^{-(l)}_{i}=\overline{e}\vphantom{e}^{k}_{(i)}\partial _{k},
 \end{equation}
where we introduced the complex orthonormal dreibein
\begin{equation}\label{4.  6}
e^{k}_{(i)}=\varkappa \delta ^{k}_{i}+iR^{-1}\varepsilon _{ikl}x^{l} \quad,\quad
 e^{l}_{(i)} e_{(k)l}=\delta_{ik}.
\end{equation}
Combining~(\ref{4.  5a}) and~(\ref{4.  7}) yields
$$\begin{eqalign}
-4R^{-2} C_{2}^{+(l)}=g^{i k}\partial_{i} \partial_{k}+
e^{(i)l}_{\ \ \ ,  k}e^{k}_{(i)}\partial_{l} \\
-4R^{-2} C_{2}^{-(l)}=g^{ik}\partial_{i} \partial_{k}+\overline{e}
\vphantom{e}^{(i)l}_{\ \ \ \ ,  k}
 \overline{e}\vphantom{e}^{k}_{(i)}\partial_{l}.
 \end{eqalign}$$
To evaluate of the second term in the right hand side let us write
\begin{equation}\label{4. 10}
e^{(i)l}_{ \ \ \ ,  k}e^{k}_{(i)}=(e^{(i)l}_{\ \ \ \ ;k}-
\Gamma^{l}_{k m}e^{(i)m})e^{k}_{(i)}=
G^{i}_{\ ni}e^{(n) l}-
\Gamma^{l}_{k m}g^{k m},
\end{equation}
where the Ricci rotational coefficients
 $G_{i k l}=
e_{(i )m ;n} e^{m}_{(k )} e^{n}_{(l )}.  $
are introduced. In our case they equal to
\begin{equation}\label{4.  9}
G_{ikl}=iR^{-1}\varepsilon _{ikl}.
\end{equation}
 Using~(\ref{4.  9}) and the known general formula for the rolling up
 of the Christoffel symbols
\begin{equation}\label{4. 12}
\Gamma^{l}_{k m}g^{k m}=-g^{-1/2}\partial_{k}
(g^{1/2}g^{lk}),
\end{equation}
where $g=\det g_{ik}$, we obtain
the  Casimir operators  of the scalar representation:
\begin{equation}\label{4. 11}
-4R^{-2}C^{+(l)}_{2}=-4R^{2}C^{-(l)}_{2}=
g^{-1/2}\partial _{i}\{ g^{1/2}g^{i k}
\partial_{k}\} \equiv\Delta.
\end{equation}
This result does not depend(hold)  on the manifest
form of vierbeins, but only on correctness of~(\ref{4. 9}).
We choose the representation with the weights $j_{+}=j_{-}=\frac{1}{2}
 (i\omega R-1)$, where $\omega\in {\Bbb R}$.  Then by  calculating of the
eigenvalues of the Casimir operators we obtain the Laplace equation:
$$ (\Delta  +\omega^{2}+R^{-2})\psi  =0  $$
which describes the field of a mass $\omega$ in the $H^3$ space.
In the space ${\Bbb R}^{1}\otimes H^3$ the quantity $\omega$ play a part
of the frequency: $\psi\sim e^{i\omega t}$ and we obtain the Klein-Gordon
equation
\begin{equation}\label{4. 11a} (\Box+
R^{-2})\psi=0 \quad,\quad \Box\equiv (-\det g_{\mu\nu})^{-1/2}
\partial_{\mu}((-\det g_{\mu\nu})^{1/2}g^{\mu\nu}\partial_{\nu})
 =-\frac{\partial^2}{\partial t^2}+\Delta .
\end{equation}
for the conformally-coupled scalar field over ${\Bbb R}^{1}\otimes H^3$.

\section {Spin-1/2 and 1 particles; Weyl and Maxwell   equations
\label{III.3}}
Let us take a certain matrix representation of the rotation group with
generators ${\bf X}^{(s)} $ which corresponds to the spin $s$.
Then the generators
 $${\bf P}^{(s)}=\frac{i}{R}{\bf X}^{(s)}$$
obey the commutation relations~(\ref{4. 3/1}),(\ref{4.
3/3}); the corresponding representation of the $SO(3,1)$ is ${\sf L}^{(s,0)}$.
 For the spin~1/2  representation
\begin{eqalignno}\label{4.  13}
{\bf X}^{(s)}=-\frac{i\bm{\sigma}}{2} \nonumber \\
X^{(s)i}X^{(s)k}=
-\frac{1}{4} \delta^{ik}+\frac{1}{2} \varepsilon^{ikl} X^{(s)l}.
\end{eqalignno}
 For the spin~1  representation
\begin{eqalignno}
(X^{(s)}_{i})_{kl}=-\varepsilon_{ikl} \nonumber\label{4.  14a} \\
(X^{(s)}_{i}X^{(s)}_{k})_{mn}=-\delta_{ik}\delta_{mn}+\delta_{in}\delta_{km}.
\label{4.  14b}
\end{eqalignno}
Now, let us consider  the representation with the generators
${\bf P}^{(l)}+{\bf P}^{(s)}$ and ${\bf X}^{(l)}+{\bf X}^{(s)}$.
The corresponding Casimir operators are
\begin{equation}\label{4. 15}
 \begin{eqalign}
-4R^{-2}C_{2}^{+}=\Delta +\frac{4s}{R}A+\frac{4}{R^2}s(s+1) \\
 -4R^{-2}C_{2}^{-}=\Delta,
\end{eqalign}
\end{equation}
where we use~(\ref{4. 11}) and designate the spinor wave operator on $H^3$
as
$$A =is^{-1} X^{(s)i}e_{(i)}^{k}\partial_{k}. $$
Let us square the above expression:
 $$-s^{2}A^{2}=X^{(s)}_{i}X^{(s)}_{k}e_{(i)}^{m}e_{(k)}^{n}(\partial_{m}
 \partial_{n}-\Gamma_{mn}^{l}\partial_{l})+\frac{s}{R}A,$$
where we expressed the derivatives of dreibein through
$G_{ikl}$ and $\Gamma^{l}_{mn}$ by analogy with~(\ref{4. 10}).
Using~(\ref{4. 14b}) and~(\ref{4. 12}) we obtain for the case $s=1/2$:
\begin{equation}\label{4. 16}
 A^{2}=\Delta-\frac{2A}{R}.
\end{equation}
From the other hand, using~(\ref{4. 3/2}) and~(\ref{4. 9}) gives
 $$-s^{2}A^{2}=X^{(s)}_{k}X^{(s)}_{i}e_{(i)}^{m}\partial_{m}(e_{(k)}^{n}
 \partial_{n})+\frac{2s}{R}A.$$
Substituting~(\ref{4. 13}) in the above expression, we obtain for the spin~1
 in the component-wise form
$$-(A^{2}\psi)_{i}=-\Delta\psi_{i}+e_{(i)}^{m}\partial_{m}
(e_{(k)}^{n}\partial_{n}\psi^{k})+\frac{2}{R}(A\psi)_{i}.$$
As  both $\Delta$ and $A$ should have fixed eigenvalues in the
irreducible representations, then~(\ref{4. 16}) holds as before
and the gauge condition
\begin{equation}\label{4. 17}
 e_{(i)}^{k}\partial_{k}\psi^{i}=0.
\end{equation}
should be
 satisfied. Using~(\ref{4. 8}),(\ref{4. 8a}),(\ref{4. 8b}),(\ref{4. 15})
and~(\ref{4. 16}) we obtain that our representations has the weights
$$j_{+}=-\frac{i\omega
R+s+1}{2} \quad,\quad j_{-}=\frac{-i\omega R+s-1}{2},$$
where $\omega\in {\Bbb R}$ is a frequency. The eigenvalues of the wave
operators are
$$\begin{eqalign}
 A\psi=\left( i\omega-\frac{s+1}{R}\right)\psi \\
 \left( \Delta-\frac{(i\omega R-s)^{2}-1}{R^2} \right) \psi=0.
\end{eqalign}$$
Proceeding to the space ${\Bbb R}^{1}\otimes H^3$ we obtain that
both for spin~1/2 particles and for spin~1 particles the equation
\begin{equation}\label{4.  18}
(\partial_{t}-is^{-1}X^{(s)i}e_{(i)}^{k}\partial_{k}-\frac{s+1}{R})\psi =0,
\end{equation}
is correct  as well as the gauge condition~(\ref{4. 17})  for spin~1
particles is. The obtained equations are the Weyl equations for the neutrino
and the Maxwell equations for the photon. If we start from the matrix
representations ${\sf L}^{(0,s)}$ then we came to the equation
\begin{equation}\label{4.  19}
(\partial_{t}+is^{-1}X^{(s)i}e_{(i)}^{k}\partial_{k}+\frac{s+1}{R})\psi =0
\end{equation}
which corresponds to the antineutrino and photon.

\section{Explicit solutions of the wave equations \label{III.4}}
Let $\bf k$ is a unit  3-vector: ${\bf k}^{2}=1$; let us consider the
$2s+1$-component quantities $u_{s}^{\pm}({\bf k})$ which obey the condition
$$(is^{-1}{\bf kX}^{(s)}\pm 1)u_{s}^{\pm}({\bf k})=0.$$
In the case $s=0\ u({\bf k})\equiv 1$. In the case of spin~1 we also
impose the one more condition, thus
\begin{equation}\label{4. 19a}
{\bf u}^{2}={\bf ku}=0.
\end{equation}
The explicit form of the $u_{s}^{\pm}({\bf k})$ see in Appendix.
Now, let us construct  the functions
$$\varphi^{(s)\pm}_{\bf k\omega}({\bf x})=\left( \varkappa-\frac{{\bf
kx}}{R} \right)^{i\omega R-s-1} u_{s}^{\pm}({\bf k}).$$
One can omit the symbol $\pm$ at the functions
$\varphi^{(0)\pm}_{\bf k\omega}({\bf x})$.
It is easy to show that the functions
\begin{equation}\label{4. 21}
\psi({\bf x},t)=e^{i\omega t}\varphi^{(s)+}_{\bf k\omega}({\bf x})
\end{equation}
obey Eq.~(\ref{4. 11a}) for $s=0$,~(\ref{4. 18}) for $s=1/2$
and~(\ref{4. 17}),(\ref{4. 18}) for $s=1$. In  two last cases it is
necessary to use the equality
$$(\varepsilon_{ijl}X^{(s)}_{i}k^{j}x^{l}+s{\bf kx} \pm i{\bf
X}^{(s)}{\bf x}) u_{s}^{\pm}({\bf k})=0,$$
that we prove easy taking into account  the explicit form of the
generators and the condition~(\ref{4. 19a}).
The functions
\begin{equation}\label{4. 23}
\psi({\bf x},t)=e^{i\omega t}\varphi^{(s)-}_{\bf k\omega}({\bf x})
\end{equation}
obey the Eq.~(\ref{4. 19})
\begin{equation}\label{4.  22}
(\partial_{t}+is^{-1}X^{(s)i}\overline{e}
\vphantom{e}_{(i)}^{k}\partial_{k}-\frac{s+1}{R})\psi =0
\end{equation}
for the antiparticles in the complex-conjugate dreibein.
To prove that in the spin~1 case the solution~(\ref{4. 21}) really
describes the "plane" electromagnetic wave let us divide $\psi$ onto
the electrical and magnetic fields:
$$\bm{ \psi}={\bf E}+i{\bf H}.  $$
As $\bm{\psi}^{2}={\bf k}\bm{\psi}=0$ then
 ${\bf E}^{2}={\bf H}^{2}$ and thus ${\bf E}, {\bf H}$ and ${\bf k}$
are orthogonal to each other, i.e. our solution really
describes the "plane" electromagnetic wave.
Functions $\varphi^{(s)\pm}_{\bf k\omega}({\bf x})$ are
analogous to
the "plane waves" on the de Sitter space~\cite{24,II}.  With
 $R\rightarrow\infty$ functions~(\ref{4. 21}) pass onto the usual plane
waves over the Minkowski space:
$$\lim_{R\rightarrow\infty}e^{i\omega t}\varphi^{(s)\pm}_{\bf k\omega}({\bf
x})=\exp (ip_{\mu}x^{\mu})u_{s}^{\pm}({\bf p}),$$
where $p^{\mu}=(\omega,\omega {\bf k})$.
The solutions we have obtained  are much simpler than the solutions of the
spin~0 and~1/2 wave equations in the Einstein space obtained for the
spin~1/2 by the method of the separation of variables in~\cite{15}
and for the spin~0,1/2 and~1  using  some group theoretical methods
in~\cite{42}.

\section{Comparison with the covariant approach\label{III.5}}
Let us complete the dreibein~(\ref{4. 6}) to the vierbein by
\begin{equation}\label{4. 24}
e^{0}_{(i)}=e^{i}_{(0)}=0 \quad,\quad  e^{0}_{(0)}=1,
\end{equation}
then  $e^{\mu}_{(\rho )}e^{(\rho )\nu}=g^{\mu \nu}$. Substituting
this in the massless general-covariant Dirac equation one can obtain
\begin{equation}\label{4.  25}
\begin{eqalign}
i\gamma ^{\mu}e_{(\mu)}^{\nu} {\cal D}_{\nu} \psi =0 \\
{\cal D}_{\nu}=\partial _{\nu}-
\frac{1}{4}G_{\rho\sigma \mu}\gamma^{\rho}\gamma^{\sigma}e^{(\mu)}_{\nu}.
\end{eqalign}
\end{equation}
The possibility of  using the complex vierbein~(\ref{4. 6}),(\ref{4. 24})
in the Weyl and Dirac equations is proved by the fact that the coincidence of
the vector and spinor $SO(3)$ transformations
\begin{equation}\label{4. 26}
L_{ik}=\delta_{ik}\cos\vartheta+\varepsilon_{ikj}l^{j}\sin\vartheta+
2l^{i}l^{k}\sin^{2}\vartheta/2 \Leftrightarrow
U(g)=\exp(-i\vartheta\bm{\sigma}{\bf l}/2),
\end{equation}
where ${\bf l}^{2}=1$ so that
$$L_{ik}(g)U(g)\sigma_{k}U^{-1}(g)=\sigma_{i} \quad,\quad g\in SO(3),$$
may be formally continued to the area of complex rotation parameters.
Since the complex conjugation is nowhere used
in the~(\ref{4. 25}) and (\ref{4. 26}), then
Eq.~(\ref{4. 25}) is invariant under the complex rotations of
the vierbein. However, the vierbein~(\ref{4. 6}),(\ref{4. 24})
may be transformed into the real vierbein
$$e_{(i)}^{k}=\delta_{ik}+\frac{x^{i}x^{k}}{R^{2}(\varkappa+1)} \quad,\quad
e^{0}_{(i)}=e^{i}_{(0)}=0 \quad,\quad  e^{0}_{(0)}=1$$
with the help of rotations around the unit vector
 ${\bf l}={\bf x}/r$ on the imaginary angle $\vartheta$,
$\cos\vartheta=\varkappa$.

Splitting~(\ref{4. 25})\footnote{
Also it is convenient to pass to the signature $(+1,-1,-1,-1)$
temporarily, in order to have the possibility of  using  the usual
$\gamma$-matrices in the standard representation.},
we obtain two equations which coincides with~(\ref{4. 18}) and~(\ref{4. 19})
for the spin~1/2. By this the general-covariance of the mentioned equations
is proven also. Let us use this general-covariance for the transformation
of the functions~(\ref{4. 23}) from the vierbein
$\overline{e}\vphantom{e}_{(\mu)}^{\nu}$ to the vierbein $e_{(\mu)}^{\nu}$.
The corresponding three-dimensional transformation is the rotation of the
form~(\ref{4. 26}) around ${\bf l}={\bf x}/r$  on the angle $\vartheta$,
$\sin\vartheta/2=ir/R$. Corresponding spinor transformation is
$$V({\bf x})=\varkappa-\frac{\bm{\sigma}{\bf x}}{R}.$$
Then the functions
\begin{equation}\label{4. 27}
\psi({\bf x},t)=e^{i\omega t}V({\bf x})
\varphi^{(1/2)-}_{\bf k\omega}({\bf x})
\end{equation}
obey Eq.~(\ref{4. 19}) for the spin~1/2.

Now, let us consider the spin 1 particles. Let us introduce the vierbein
tensor of the electromagnetic field
$$\Phi_{\mu \nu}=e_{(\mu)}^{\rho}e_{(\nu)}^{\sigma}F_{\rho\sigma}, $$
where
$F_{\rho\sigma}$ is the  usual tensor of the electromagnetic field. By
expressing the covariant derivatives of $F_{\mu\nu}$ in terms of vierbein
derivatives of $\Phi_{\rho\sigma}$ we obtain the Maxwell equations
\begin{equation}\label{4.  28}
\begin{eqalign}
e_{(\nu)}^{\sigma}\partial_{\sigma}
\Phi^{\mu\nu}-G^{\mu\nu\rho}\Phi_{\nu\rho}-
\Phi^{\sigma \mu}G_{\sigma \varkappa}^{\ \ \varkappa}=0 \\
\varepsilon^{\rho\mu\nu \varkappa}(e_{(\varkappa)}^{\sigma}\partial_{\sigma}
\Phi_{\mu\nu}+
2\Phi^{\sigma}_{\ \nu}G_{\sigma \mu\varkappa})=0
\end{eqalign}
\end{equation}
Substituting~(\ref{4. 9}) into~(\ref{4. 28}) and introducing the complex
vector of the electromagnetic field
$$\psi_{i}=\Phi_{i0}+\frac{i}{2}\varepsilon_{ikl}\Phi_{kl}, $$
we obtain that~(\ref{4. 28}) coincides with~(\ref{4. 17}) and~(\ref{4. 18})
for the spin~1 case.

\section{Propagators and two-point functions\label{III.6}}
The functions $\varphi_{\bf k\omega}^{(0)}({\bf x})$
are the generalized coherent states for the group $SO(3,1)$~\cite{coher1};
from here its nontrivial transformation properties follow.
Let us juxtapose to each element
 $g\in SO(3,1)$ the matrix $L^{\alpha}_{\ \beta}(g)$ of
the orthogonal transformations of the space $H^3$.
Then we can define the
action of the group $SO(3,1)$ over the space
$H^3$ and over the null 4-vectors $n^\alpha$, $n^{\alpha}n_{\alpha}=0$:
\begin{equation}\label{4. 29}
x^{\alpha}\longmapsto x^{\alpha}_{g}=L^{\alpha}_{\ \beta}(g)x^{\beta}
\quad,\quad
n^{\alpha}\longmapsto n^{\alpha}_{g}=L^{\alpha}_{\ \beta}(g)n^{\beta}.
\end{equation}
Then  the
unit 3-vector $\bf k$ can be expressed through $n^{\alpha}$:
${\bf k}={\bf n}/n^{4}$, which gives us the action
of the group  $SO(3,1)$ on the unit
3-vectors: ${\bf k}\mapsto {\bf k}_{g}$. Then it is easy to show that the
functions $\varphi_{\bf k\omega}^{(0)}({\bf x})$
transform under the transformations~(\ref{4. 29}) as follows:
\begin{equation}\label{4. 30}
\varphi_{\bf k\omega}^{(0)}({\bf x}_{g})=
(L_{4}^{\ 4}(g)+L_{4}^{\ i}(g)k^{i})^{-i\omega R-1}
\varphi_{{\bf k}'\omega}^{(0)}({\bf x}),
\end{equation}
where ${\bf k}'={\bf k}_{g^{-1}}$ and $\lambda\in {\Bbb C}$.
The correctness of the
above expression under the $g\in SO(3)$ is obvious. It is necessary
to prove this only for the transformations
$g=g_{\bm{\scriptstyle\xi}}$, which is given by the matrices
\begin{equation}\label{4. 31}
L_{4}^{\ 4}(g)=(1+{\bm{\xi}}^{2})^{1/2}
\quad,\quad L^{i}_{\ 4}(g)=L^{4i}(g)=\xi^{i}
\quad,\quad L_{ik}=\delta_{ik}+
\frac{\xi^{i}\xi^{k}}{1+\sqrt{1+{\bm{ \xi}}^{2}}}.
\end{equation}
By calculating  the Jacobian of the transformation from the
${\bf k}$ to the ${\bf k}'$ it is easy to show that the two-point function
$${\cal W}^{(0)}({\bf x},{\bf y};\omega)=
\int_{S^{2}} d^{2}k\,
\varphi_{\bf k\omega}^{(0)}({\bf x})
\varphi_{\bf k,-\omega}^{(0)}({\bf y}),$$
where $\bf x$ and $\bf y$ are the arbitrary points of the upper half of the
 space $H^3$ and $d^{2}k=dk^{1}dk^{2}/k^{3}$ is an invariant measure,
possesses the $SO(3,1)$-invariance:
\begin{equation}\label{4. 32}
{\cal W}^{(0)}({\bf x}_{g},{\bf y}_{g};\omega)=
{\cal W}^{(0)}({\bf x},{\bf y};\omega).
\end{equation}
In fact, ${\cal W}^{(0)}({\bf x}_{g},{\bf y}_{g};\omega)$
is the scalar product of the coherent states  corresponding
to the points $\bf x$ and $\bf y$. Then the equality
$${\cal W}^{(0)}({\bf x},{\bf y};\omega)=F\left(
-\frac{i\omega R-1}{2},
\frac{i\omega R+1}{2},\frac{3}{2};
1-\left(\frac{x_{\alpha}y^{\alpha}}{R}\right)^{2}\right).$$
is correct~\cite{coher1}.

The above considerations may be generalized to the cases of
spin~1/2 and~1 too.
Let us construct the $(2s+1)\times (2s+1)$-matrices
$$\Sigma ({\bf k};s)=u_{s}^{+}({\bf k})\otimes u_{s}^{+\dagger}({\bf k}).$$
We can show (see Appendix)  the validity of
\begin{equation}\label{4. 33}
U_{s}(g)\Sigma({\bf k};s)U_{s}^{\dagger}(g)= \left( \frac{n^4_g}{n^4}
\right)^{2s} \Sigma ({\bf k}_{g};s),
\end{equation}
where $U_{s}(g)$ is the matrix of the representation ${\sf L}^{(s,0)}$
which corresponds to the transformation $g\in SO(3,1)$.
Then, by the way completely analogous to that for the spin zero particles
and using~(\ref{4.  33}) we can obtain that the two-point functions
$${\cal W}^{(s)}({\bf x},{\bf y};\omega)= \int_{S^{2}} d^{2}k\, \varphi_{\bf
k\omega}^{(s)+}({\bf x}) \otimes \left( \varphi_{\bf k\omega}^{(s)+}({\bf
y})\right)^{\dagger}$$
possess the $SO(3,1)$-invariance:
$${\cal W}^{(s)}({\bf x}_{g},{\bf y}_{g};\omega)= U_{s}(g){\cal W}^{(s)}
({\bf x},{\bf y};\omega)U^{\dagger}_{s}(g).$$
Now, to clarify of the physical meaning of the constructed
two-point functions
let us construct now the secondly quantized massless spin 0, 1/2 and~1 fields:
$$\Psi^{(s)}(x)=\int_{0}^{\infty} \frac{\omega
d\omega}{(2\pi)^{3/2}}\int_{S^2}d^{2}k\, \left( e^{i\omega
x^0}\varphi^{(s)+}_{\bf k\omega}({\bf x}) a({\bf p};s)+ e^{-i\omega
x^0}\varphi^{(s)+}_{\bf k,-\omega}({\bf x}) a^{\dagger}({\bf p};s)\right), $$
where the bosonic and fermionic creation-destruction operators
$$[a_{A}({\bf p};s),a^{\dagger}_{B}({\bf p}';s)]_{\pm}=
p^{0}\delta^{3}({\bf p}-{\bf p}')\delta_{AB}$$
are introduced and $A,B=1,\ldots,2s+1$.
The choice of a measure is correct because
$$\omega\, d\omega \, d^{2}k=\frac{d^{3}{\bf p}}{p^0}.$$
Then the corresponding propagators
\begin{eqnarray}
\Delta^{(s)}_{AB}(x,y)\equiv
[\Psi^{(s)}_{A}(x),\Psi^{(s)\dagger}_{B}(y)]_{\pm} = \nonumber \\
=\int_{0}^{\infty}\frac{\omega d\omega}{(2\pi)^3}\left(
e^{i\omega(x^{0}-y^{0})}{\cal W}^{(s)}_{AB}({\bf x},{\bf y};\omega)-
e^{-i\omega(x^{0}-y^{0})}{\cal W}^{(s)}_{AB}({\bf y},{\bf x};\omega) \right)
\nonumber
\end{eqnarray}
possess the invariance under the time translations and spatial
 $SO(3,1)$ transformations. The equality
$$\int \frac{d^{3}{\bf x}}{\varkappa}\, \varphi^{(0)}_{\bf k\omega}({\bf
x}) \varphi^{(0)}_{{\bf k}',-\omega'}({\bf x})=
(2\pi)^{3}\delta^{3}({\bf p}-{\bf p}'),$$
is correct~\cite{coher1}, where
${\bf k,k}',\omega,\omega'$ are  arbitrary.
 Then, it is easy to show that the propagator for spin zero we have
constructed  possess the reproductivity property:
$$\int \frac{d^{3}{\bf x}}{\varkappa}\,
\Delta^{(0)}(y,x)\Delta^{(0)}(x,z)=\Delta^{(0)}(y,z).$$

\section{Other reducible spaces\label{III.7}}
The passage from the space ${\Bbb R}^1 \otimes H^3$ to the Einstein space
${\Bbb R}^1 \otimes S^3$ may be performed by the change of sign of
$\eta_{44}$ and by the replacement $R\rightarrow iR$.
The symmetry group of $S^3$ is $SO(4)=SO(3)\otimes
SO(3)$; the dreibein~(\ref{4. 6}) becomes real, and therefore the operators
$\Pi_{i}^{\pm}$ becomes the generators of two subgroups of true translations
of the space $S^3$. The invariant wave operators remains unchanged to
within the replacement $R\rightarrow iR$,
but their eigenvalues no longer may be obtained by the group-theoretical way.
To within the same replacement, the explicit solutions of the wave equations
and the general-covariant equations remain unchanged, but many results
of~\sect{III.5} don't take a place since the vierbein~(\ref{4. 6}) becomes
real. The results of~\sect{III.6} don't take a place.

Now, let us consider the space with the symmetry group $ T_{1} \otimes SO (3)
$. According to classification of all spaces with four-parametric symmetry
groups given in~\cite{13}, spaces with such a symmetry group  form the 
whole class (type VIII in the mentioned classification) and they are the 
direct products of a temporal axes and three-dimensional spaces with the 
 metric depending on the constant parameters $K_{ij} = K_{ji}$ and $D $. The  
scalar representation generators of their symmetry group are 
$$\Pi^{(l)}_{\mu} = b_{(\mu)}^{\nu} \partial_ {\nu}, $$
where $b_{(\mu)}^{\nu}$ are four vectors, the nonzero component
of which are
$$\begin{eqalign}
b_{(0)0}=-1 \quad ; \quad b_{(1)1}=\cos(y/R) \quad ; \quad b_{(2)2}=1 \\
b_{(3)3}=\cos(x/R)\cos(y/R) \quad ;\quad b_{(1)3}=\cos (x/R)\sin (y/R) \\
b_{(2)3}=-\sin  (x/R) \quad ; \quad b_{(3)1}=-\sin (y/R).
\end{eqalign}$$
It is necessary that these vectors  compose an orthonormal
 vierbein for the group theoretical approach has meaning.
It follows from here that  the nonzero components of the metric
 tensor are
$$g_{11}=g_{22}=g_{33}=1 \quad ;\quad g_{23}=-\sin (x/R).  $$
It is easy to prove that this metric really belongs to the
specified class and has nonzero parameters $K_ {11} = 2 \ , \ K_{33} =D=1$.
Now, the Ricci rotational coefficients are
$$G_{ikl}=R^{-1}\varepsilon_{ikl}.$$
As all the relations of~\sect{III.2},\sect{III.3},\sect{III.5} were
derived only with the use of~(\ref{4. 9})  and without using  the
explicit form of vierbein, then all the formulas (except the explicit
solutions of the wave equations) which take place in the
${\Bbb R}^1 \otimes S^3$ space, take place in the Petrov type VIII too.
Only difference is that instead  two
$SO(3)$-subgroups remains only one, but it is enough for the construction
of the invariant wave operators.

\section*{Appendix. Some properties of $u_{s}^{\pm}({\bf k})$}
Let us juxtapose to  each element $g\in SO(3,1)$ the matrix
$\left(
\begin{array}{ll}
a & b \\ c & d
\end{array}
\right)$
of the group $SL(2,{\Bbb C})\sim SO(3,1)$.
There exists its fractionally-linear transformation on the imaginary line:
$$z \mapsto z_{g}=\frac{az+b}{cz+d}.$$
Then we can determine the group $SL(2,{\Bbb C})$ representation
acting on the functions of $z$ by the following way
$$f(z)\mapsto (cz+d)^{2s}f(z_{g^{-1}}).$$
As is well known (see~\cite{49}) if $f(z)$ is the polynomial from $z$ of
 the power $2s$, then the representation is ${\sf L}^{(s,0)}$.
This means that it is possible to write down the vectors
of the representation ${\sf L}^{(s,0)}$ not in the form of the lines or
columns, but in the form of polynomials of a power $2s$.
Then one can obtain~\cite{40} that in  such a notation
$u_{s}^{+}({\bf k})$ with the arbitrary $s$ looks like
$$u_{s}^{+}({\bf k})=\left(\frac{1+k^3}{2}\right)^{s}
(1-z\overline{\rho}_{\bf k})^{2s},$$
where
$$\rho_{\bf k}=\frac{k^{1}+ik^2}{1+k^3}.$$
The quantity  $\rho_{\bf k}$ has a simple geometrical meaning. Indeed,
let us make the stereographic projection of the sphere $S^2$ where
the ends of the vectors $\bf k$ are lying, onto the plane. Here
each point $(x,y)$ of the plane is in the conformity with the complex 
quantity $x+iy$.  Then, the quantity $\rho_{\bf k}$ corresponds to the vector 
$\bf k$.  This means that the fractionally-linear action of the $SL(2,{\Bbb 
C})$ onto the $\rho_{\bf k}$ is equivalent to the projective action of the 
$SO(3,1)$ onto the $\bf k$:  $(\rho_{\bf k})_{g}= \rho_{{\bf k}'}$, where 
${\bf k}'={\bf k}_{g}$.  Thus, it is easy to show that
$$\left( U_{s}(g)u_{s}^{+}({\bf k})\right)(z)=(cz+d)^{2s}u_{s}^{+}({\bf k})
(z_{g^{-1}})=(\alpha_{\bf k}(g))^{s}u^{+}_{s}({\bf k}_{g^{-1}}) (z),$$
where
$$\alpha_{\bf k}(g)=\frac{1+k^3}{1+k^{3}_{g^{-1}}}
(a-b\rho_{\bf k})^{-2}$$
is independent on $s$ and $z$. The change of $\Sigma({\bf k};s)$ under
the transformation is determined by the $|\alpha_{\bf k}(g)|^{2}$; to
obtaining this quantity in the convenient form let us consider the case of 
spin~1/2.  Then in the usual "columnar" notation we have
$$ u^{+}({\bf k})=\sqrt{\frac{1+k^3}{2}}\left(
\begin{array}{r}
-\overline{\rho}_{\bf k} \\ 1
\end{array} \right) \quad,\quad
u^{-}({\bf k})=\sqrt{\frac{1+k^3}{2}}\left(
\begin{array}{r}
1 \\ \rho_{\bf k}
\end{array} \right).$$
It is easy to show that
$$\Sigma({\bf k};1/2)=-\frac{1}{2} \frac{\sigma_{\alpha}n^{\alpha}}{n^4},$$
where $\sigma^{\alpha}=(\bm{\sigma},1)$. As
$$U(g)\sigma_{\alpha}n^{\alpha}U^{\dagger}(g)=\sigma_{\alpha}n^{\alpha}_{g},$$
we obtain
$$|\alpha_{\bf k}(g)|^{2}=\left(\frac{n^4_g}{n^4}\right)^{2}.$$
Putting everything together gives the equality~(\ref{4. 33}) looked for.

\end{document}